\documentstyle[sprocl,epsfig]{article}
\newcommand{\lesssim}{\raisebox{0.3mm}{\em $\, <$} \hspace{-2.8mm}
\raisebox{-1.3mm}{\em $\sim \,$}}

\def\be{\begin{equation}}
\def\ee{\end{equation}}
\def\bea{\begin{eqnarray}}
\def\eea{\end{eqnarray}}

\begin{document}

\title{VARIOUS SOLUTIONS OF THE ATMOSPHERIC NEUTRINO DATA}

\author{OSAMU YASUDA}

\address{Department of Physics,
Tokyo Metropolitan University \\
Minami-Osawa, Hachioji, Tokyo 192-0397, Japan
\\E-mail: yasuda@phys.metro-u.ac.jp}


\maketitle\abstracts{Various solutions of the atmospheric neutrino
data are reviewed.  Apart from orthodox two flavor
$\nu_\mu\leftrightarrow\nu_\tau$ oscillations and three flavor
oscillations, there are still possibilities, such as four flavor
oscillations with the (2+2)- and (3+1)- schemes, a neutrino decay
scenario and decoherence, which give a good fit to the data.}

\section{Introduction}

It has been known that the atmospheric neutrino anomaly
\cite{Kamatm,Kamup,IMB,SKatm,SKup,SKs,learned,mcgrew,soudan2,macro,atmth}
can be accounted for by dominant $\nu_\mu\leftrightarrow\nu_\tau$
oscillations with almost maximal mixing,
and the zenith angle dependence of atmospheric neutrinos has
been analyzed by many theorists \cite{atmanalysis} as well as
by experimentalists \cite{Kamatm,Kamup,SKatm,SKup,SKs,soudan2,macro}.
On the other hand, the solar neutrino
observations \cite{homestake,Kamsol,SKsol,ysuzuki,sage,gallex}
and the LSND experiment \cite{lsnd,mills}
also suggest neutrino oscillations.
$\Delta m^2_{\mbox{\rm\footnotesize atm}}$,
$\Delta m^2_\odot$ and
$\Delta m^2_{\mbox{\rm{\scriptsize LSND}}}$
(the mass squared differences suggested by the atmospheric neutrino
anomaly, the solar neutrino deficit and the LSND data) have
different orders of magnitudes and there have been a lot of works
to analyze the atmospheric neutrino data from the view point
of ordinary oscillations due to mass with two, three and
four flavors as well as exotic scenarios.
In this talk I will review the status of various
scenarios which have been proposed to explain
the atmospheric neutrino problem.

\section{Neutrino oscillations due to mass}

\subsection{Neutrino oscillations with two flavors}

The most up-to-date result of the two flavor
analysis of $\nu_\mu\leftrightarrow\nu_\tau$ with 1289 day data
has been given by
McGrew \cite{mcgrew} and the allowed region
of the oscillation parameters at 90\%CL is
\begin{eqnarray}
0.88<&\sin^22\theta_{\mbox{\rm\footnotesize atm}}&\le 1\nonumber\\
1.6\times10^{-3}{\rm eV}^2<
&\Delta m_{\mbox{\rm\footnotesize atm}}^2& < 4\times10^{-3}{\rm
eV}^2.\nonumber
\end{eqnarray}
On the other hand, two flavor
analysis of $\nu_\mu\leftrightarrow\nu_s$ has been done
by the Superkamiokande group using the data of
neutral current enriched multi-ring events,
high energy partially contained events and upward going $\mu$'s,
and they have excluded the two flavor oscillation
$\nu_\mu\leftrightarrow\nu_s$ at 99\%CL \cite{SKs}.

\subsection{Neutrino oscillations with three flavors}

The flavor eigenstates are related to the mass eigenstates
by the $3\times3$ MNS mixing matrix:
\begin{eqnarray}
\left(
\begin{array}{ccc}
U_{e1} & U_{e2} &  U_{e3}\\
U_{\mu 1} & U_{\mu 2} & U_{\mu 3} \\
U_{\tau 1} & U_{\tau 2} & U_{\tau 3}
\end{array}\right)=
\left(
\begin{array}{ccc}
c_{12}c_{13} & s_{12}c_{13} &  s_{13}e^{-i\delta}\\
-s_{12}c_{23}-c_{12}s_{23}s_{13}e^{i\delta} &
c_{12}c_{23}-s_{12}s_{23}s_{13}e^{i\delta} & s_{23}c_{13}\\
s_{12}s_{23}-c_{12}c_{23}s_{13}e^{i\delta} &
-c_{12}s_{23}-s_{12}c_{23}s_{13}e^{i\delta} & c_{23}c_{13}\\
\end{array}
\right),\nonumber
\end{eqnarray}
and without loss of generality I assume $|\Delta
m_{21}^2|<|\Delta m_{32}^2|<|\Delta m_{31}^2|$
where $\Delta m^2_{ij}\equiv m^2_i-m^2_j$,
$m_j^2 (j=1,2,3)$ are the mass squared for the mass
eigenstates.  Since there
are only two independent mass squared differences, it is
impossible to account for the solar neutrino deficit, the atmospheric
neutrino anomaly and LSND (the only nontrivial possibility is to take
$\Delta m^2_{21}=\Delta m^2_{\mbox{\rm\footnotesize atm}}$ and
$\Delta m^2_{32}=\Delta m^2_{\mbox{\rm{\scriptsize LSND}}}$
and to try to explain the solar neutrino
problem with the energy independent solution; It turns out, however,
that the main oscillation channel in the atmospheric neutrinos in this
case is $\nu_\mu\leftrightarrow\nu_e$ and therefore the zenith angle
dependence of the atmospheric neutrino data cannot be explained).
So I have to give up an effort to explain LSND and I have to
take $\Delta m^2_{21}=\Delta m^2_\odot$ and
$\Delta m^2_{32}=\Delta m^2_{\mbox{\rm{\scriptsize atm}}}$.
Under the present assumption it follows
$\Delta m^2_{\rm atm}=\Delta m^2_{32}$
$\gg$$\Delta m^2_{21}=\Delta m^2_\odot$
and I have a large hierarchy between
$\Delta m^2_{21}$ and $\Delta m^2_{32}$.
If $|\Delta m^2_\odot L/4E|\ll 1$ then from a hierarchical
condition I have the oscillation probability
\begin{eqnarray}
P(\bar{\nu}_e \rightarrow \bar{\nu}_e) 
= 1-\sin^22\theta_{13}\Delta_{32},\nonumber
\end{eqnarray}
where $\Delta_{jk}\equiv
\sin^2(\Delta m_{jk}^2 L/4E)$,
so if $\Delta m^2_{\rm atm} > 2\times10^{-3}$eV$^2$
then the CHOOZ reactor data \cite{chooz} force us to have
either $\theta_{13}\simeq 0$ or $\theta_{13}\simeq \pi/2$.
On the other hand, the solar oscillation probability
in the three flavor framework 
is related to that in the two flavor case by \cite{3to2}
\begin{eqnarray}
P^{(3)}(\nu_e\rightarrow\nu_e;A(x))
= c_{13}^4P^{(2)}
(\nu_e\rightarrow\nu_e;c_{13}^2A(x))+s_{13}^4,\nonumber
\end{eqnarray}
where $A(x)$ stands for the matter effect.
To account for the solar neutrino deficit, therefore,
$|s_{13}|$ cannot be too large, so
it follows that $|\theta_{13}|\ll 1$ and the 
MNS mixing matrix $U$ becomes
\begin{eqnarray}
U \simeq
\left(
\begin{array}{ccc}
c_\odot & s_\odot &  \epsilon\\
-s_\odot/\sqrt{2}-c_\odot/\sqrt{2} &
c_\odot/\sqrt{2}-s_\odot/\sqrt{2} & 1/\sqrt{2}\\
s_\odot/\sqrt{2}-c_\odot/\sqrt{2} &
-c_\odot/\sqrt{2}-s_\odot/\sqrt{2} & 1/\sqrt{2}\\
\end{array}
\right),
\nonumber
\end{eqnarray}
which indicates that the solar neutrino problem is explained
by oscillations half of which is
$\nu_e\rightarrow\nu_\mu$ and the other is
$\nu_e\rightarrow\nu_\tau$, and that
the atmospheric neutrino anomaly is accounted for
by oscillations of almost 100\% $\nu_\mu\rightarrow\nu_\tau$
($|\epsilon|\equiv|\theta_{13}|\ll 1$).

On the other hand, if
$\Delta m^2_{\rm atm}<$$2\times10^{-3}$eV$^2$, then
$\theta_{13}$ can be relatively large (This
possibility gives a bad fit to the atmospheric neutrino data
but is not excluded at $4\sigma$CL yet).
From the combined three flavor analysis of the
Superkamiokande atmospheric neutrino data
with the CHOOZ data,
it has been shown \cite{flmm,gmpv}
that $|\theta_{13}|\lesssim\pi/12$ is allowed at 99\%CL.
Hence the probability
$$P(\nu_\mu \rightarrow \nu_e) = s^2_{23}\sin^22\theta_{13}
\Delta_{32}$$
of appearance of $\nu_e$ can be relatively large
and there is a chance in long baseline experiments to
observe $\nu_e$ in this case.

\subsection{Neutrino oscillations with four flavors}

\begin{figure}
\begin{center}
\vglue -0.5cm \hglue -0.5cm
\epsfig{file=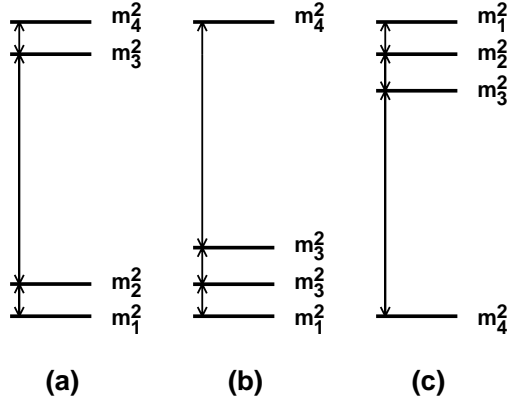,width=6cm}
\vglue 0.5cm
\caption{Mass patterns of four neutrino schemes.
(a) corresponds to (2+2)-scheme, where either
($|\Delta m^2_{21}|=\Delta m^2_\odot$,
$|\Delta m^2_{43}|=\Delta m^2_{\mbox{\rm\footnotesize atm}}$)
or
($|\Delta m^2_{43}|=\Delta m^2_\odot$,
$|\Delta m^2_{21}|=\Delta m^2_{\mbox{\rm\footnotesize atm}}$).
(b) and (c) are (3+1)-scheme, where
$|\Delta m^2_{41}|=\Delta m^2_{\mbox{\rm{\scriptsize LSND}}}$ and
either
($|\Delta m^2_{21}|=\Delta m^2_\odot$,
$|\Delta m^2_{32}|=\Delta m^2_{\mbox{\rm\footnotesize atm}}$)
or
($|\Delta m^2_{32}|=\Delta m^2_\odot$,
$|\Delta m^2_{21}|=\Delta m^2_{\mbox{\rm\footnotesize atm}}$) is satisfied.}
\label{fig:pattern}
\end{center}
\end{figure}

To explain the solar, atmospheric and LSND data within the framework
of neutrino oscillations, it is necessary to have at least four kinds of
neutrinos.
In the case of four neutrino schemes there are two distinct types of
mass patterns.  One is the so-called (2+2)-scheme
(Fig. \ref{fig:pattern}(a)) and the other is the (3+1)-scheme
(Fig. \ref{fig:pattern}(b) or (c)).  Depending on the type of the two
schemes, phenomenology is different.

The atmospheric neutrino data were analyzed by Refs.~\cite{y1,flm}
with the (2+2)-scheme.  Here I assume the mass pattern in
Fig. \ref{fig:pattern}(a) with $\Delta m^2_{21}=\Delta m^2_\odot$
and $\Delta m^2_{43}=\Delta m^2_{\mbox{\rm\footnotesize atm}}$.
I also assume $U_{e3}=U_{e4}=0$,
which is justified from the Bugey reactor
constraint $|U_{e3}|^2+|U_{e4}|^2\ll 1$,
and $\Delta m^2_\odot=0$,
since $|\Delta m^2_\odot L/2E|\ll 1$ in the atmospheric neutrino oscillations.
I take the reference value
$\Delta m^2_{\mbox{\rm{\scriptsize LSND}}}=0.3$ eV$^2$
so that
the result with large $|U_{\mu3}|^2+|U_{\mu4}|^2$ do not contradict
with the CDHSW constraint
\begin{eqnarray}
1-P(\nu_\mu\rightarrow\nu_\mu)&=&
4(|U_{\mu3}|^2+|U_{\mu4}|^2)(1-|U_{\mu3}|^2-|U_{\mu4}|^2)
\Delta_{32}\nonumber\\
&\le&\sin^22\theta_{\mbox{\rm\scriptsize CDHSW}}
(\Delta m^2_{32})\Delta_{32},\nonumber
\end{eqnarray}
where $\sin^22\theta_{\mbox{\rm\scriptsize CDHSW}}(\Delta m^2)$
stand for the value of the boundary of the excluded region of
CDHSW \cite{cdhsw} in the two flavor analysis as a function
of $\Delta m^2$.
With these assumptions,
$\nu_e$ decouples from other three neutrinos,
and the problem is reduced to the three flavor neutrino
analysis among $\nu_\mu$, $\nu_\tau$, $\nu_s$ and the reduced
MNS matrix is
\begin{eqnarray}
\widetilde U&\equiv&\left(
\begin{array}{ccc}
 U_{\mu 2} & U_{\mu 3}&U_{\mu 4} \\
 U_{\tau 2} & U_{\tau 3}&U_{\tau 4} \\
 U_{s2} &  U_{s3}&U_{s4} 
\end{array}\right)
=e^{i({\pi \over 2}-\theta_{34})\lambda_7}
D^{-1} e^{i\theta_{24}\lambda_5} D~
e^{i(\theta_{23}-{\pi \over 2})\lambda_2},\nonumber
\end{eqnarray}
with $D\equiv{\rm diag}\left(e^{i\delta_1/2},1,e^{-i\delta_1/2} \right)$
($\lambda_j$ are the $3\times 3$ Gell-Mann matrices)
is the reduced $3\times 3$ MNS matrix.
This MNS matrix $\widetilde U$
is obtained by substitution
$\theta_{12}\rightarrow\theta_{23}-\pi/2$,
$\theta_{13}\rightarrow\theta_{24}$,
$\theta_{12}\rightarrow\pi/2-\theta_{34}$,
$\delta\rightarrow\delta_1$
in the standard parametrization in Ref.~\cite{pdg}.
$\theta_{34}$ corresponds to the
mixing of $\nu_\mu\leftrightarrow\nu_\tau$ and
$\nu_\mu\leftrightarrow\nu_s$, while $\theta_{23}$
is the mixing of the contribution of
$\displaystyle\sin^2(\Delta m^2_{\mbox{\rm\footnotesize atm}}
L/4E)$
and
$\displaystyle\sin^2(\Delta m^2_{\mbox{\rm{\scriptsize LSND}}}
L/4E)$
in the oscillation probability.  The allowed region at 90\%CL
of the atmospheric neutrino data is roughly given by \cite{y1}
$30^\circ\lesssim\theta_{24}\lesssim 55^\circ$,
$0\le\theta_{23}\lesssim30^\circ$,
$-90^\circ\theta_{23}\lesssim90^\circ$.
The reasons that the (2+2)-scheme
is consistent with the recent 
Superkamiokande data are because both
solar and atmospheric neutrinos have hybrid
of active and sterile oscillations in this scheme and because
there is a constant term in the surviving probability
\begin{eqnarray}
P(\nu_\mu\rightarrow\nu_\mu)=1-
4|U_{\mu3}|^2|U_{\mu4}|^2\Delta_{43}
-2(|U_{\mu3}|^2+|U_{\mu4}|^2)(1-|U_{\mu3}|^2-|U_{\mu4}|^2)\nonumber
\end{eqnarray}
due to nonvanishing contribution of $\sin^2
(\Delta m^2_{\mbox{\rm{\scriptsize LSND}}}L/4E)$, where
I have averaged
over rapid oscillations: $\sin^2
({\Delta m^2_{\mbox{\rm{\scriptsize LSND}}}L/4E})\rightarrow1/2$.

On the other hand, 
it has been shown in Refs.~\cite{oy,bgg} using older data of LSND
\cite{lsnd} that the (3+1)-scheme is inconsistent with the Bugey
reactor data \cite{bugey} and the CDHSW disappearance experiment \cite{cdhsw}
of $\nu_\mu$.  However, in the final result the allowed region has shifted
to the lower value of $\sin^22\theta$ and it was shown \cite{bklw}
that there are four isolated regions
$\Delta m^2_{\mbox{\rm{\scriptsize LSND}}}\simeq$0.3, 0.9, 1.7, 6.0 eV$^2$
which satisfy both the constraints of Bugey and CDHSW and
the LSND data at 99\%CL.
The case of $\Delta m^2_{\mbox{\rm{\scriptsize LSND}}}$=0.3 eV$^2$
turns out to be excluded by the Superkamiokande atmospheric neutrino
data at 6.9$\sigma$CL \cite{y2}.  For the other three values of
$\Delta m^2_{\mbox{\rm{\scriptsize LSND}}}$, I have
$U_{e4}\simeq U_{\mu4}\simeq0$ and this case is
reduced to the analysis in the (2+2)-scheme with $\theta_{23}=0$.
The allowed region at 90\%CL is given roughly by
$-\pi/4\lesssim\theta_{34}\lesssim\pi/4$,
$0.8\lesssim\sin^22\theta_{24}\le1$,
where $\theta_{34}$ and $\theta_{24}$ stand for
the mixing of $\nu_\mu\leftrightarrow\nu_\tau$ and
$\nu_\mu\leftrightarrow\nu_s$ and the mixing of atmospheric
neutrino oscillations, respectively.

\section{Exotic solutions}

\begin{figure}
\begin{center}
\vglue -0.5cm
\hglue -0.5cm
\includegraphics[width=.6\textwidth]{exotic.epsi}
\vglue -0.3cm
\caption{$\chi^2$ of the SK atmospheric neutrino data
as a function of index $n$ ($1-P_{\mu\mu}=\sin^22\theta
\sin^2({\mbox{\rm const}}E^nL)$). $n = -1$ corresponds to
ordinary oscillations due to mass.}
\label{fig:enn}
\end{center}
\begin{center}
\vglue 1.0cm 
\hglue -0.5cm
\epsfig{file=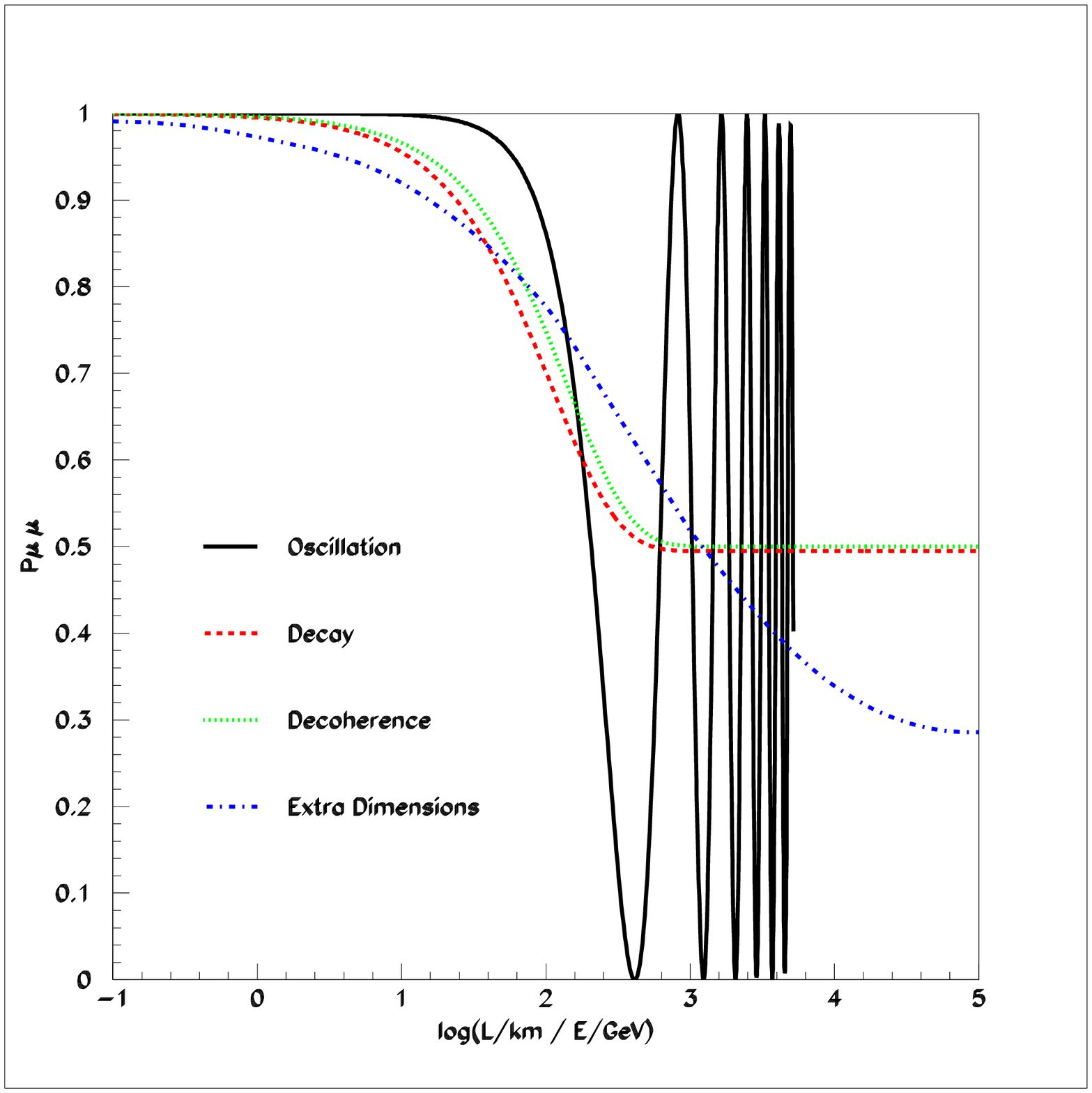,width=7.5cm}
\vglue 0.2cm 
\caption{Behaviors of the surviving probability
$P(\nu_\mu\rightarrow\nu_\mu)$ as a function of $L/E$
for scenarios of oscillations, decay, decoherence and extra dimensions.
}
\label{fig:p}
\end{center}
\end{figure}

Apart from ordinary oscillations due to mass, several possibilities
have been proposed which predict different behaviors of the
oscillation probability as a function of the neutrino energy.  Those
include violation of the equivalence principle \cite{vep}, violation
of the Lorentz invariance \cite{lorentz},
presence of torsion \cite{torsion},
flavor changing neutral
current interactions \cite{FCNC}, neutrino decays \cite{decaya,decayb}
decoherence of the neutrino beam \cite{lisi}, large extra dimensions
\cite{extra}, etc.  As in the case of test of sterile oscillations,
the zenith angle dependence (or the up-down asymmetry) of the high
energy atmospheric neutrino data give strong constraints on these
exotic scenarios.  In the case of violation of the equivalence
principle or the Lorentz invariance, the $\nu_\mu$ disappearance
probability $P_{\mu\mu}\equiv P(\nu_\mu\rightarrow\nu_\mu;L)$ is given
by
\begin{eqnarray}
P_{\mu\mu}=1-\sin^22\theta\sin^2\left({\mbox{\rm const}}\cdot EL \right)
\nonumber
\end{eqnarray}
and in the case of flavor changing neutral current interactions
\begin{eqnarray}
P_{\mu\mu}=1-\sin^22\theta\sin^2\left({\mbox{\rm const}}\cdot L \right).
\nonumber
\end{eqnarray}
Both possibilities are strongly disfavored (See Fig. \ref{fig:enn}
which is taken from Ref.~\cite{learned}).

In the case of neutrino decays, which were originally
introduced to try to explain the solar, atmospheric neutrinos
and LSND within the three flavor framework with
two oscillation parameters $\Delta m^2_{21}$, $\Delta m^2_{32}$
and one neutrino decay constant $\alpha$, the 
disappearance probability is
\begin{eqnarray}
P_{\mu\mu}=\sin^4\theta+\cos^4\theta\exp(-\alpha L/E)
+{1 \over 2}\sin^22\theta\exp(-\alpha L/2E)\cos
\left(\Delta m^2L/2E \right)
\nonumber
\end{eqnarray}
which has the following two extreme cases:
\begin{eqnarray}
P_{\mu\mu}&=&\sin^4\theta+\cos^4\theta\exp(-\alpha L/E)
\qquad\qquad\Delta m^2\rightarrow\infty~~{\mbox{\rm (case A)}},\nonumber\\
P_{\mu\mu}&=&\left[\sin^2\theta+\cos^2\theta\exp(-\alpha L/2E)\right]^2
\qquad\,\Delta m^2\rightarrow 0~~~\,{\mbox{\rm (case B)}}.\nonumber
\end{eqnarray}
If the case A gave a good fit to the data then
it would be possible to account for the solar neutrino deficit,
the atmospheric neutrino anomaly and the LSND data within
the three flavor framework by putting
$\Delta m^2_{21}=\Delta m^2_\odot$,
$\Delta m^2_{32}=\Delta m^2_{\mbox{\rm{\scriptsize LSND}}}$,
$\alpha=\Delta m^2_{\mbox{\rm\footnotesize atm}}$,
but unfortunately it is not the case.
It has been shown that the case A gives a bad fit \cite{decaya}
but the case B
gives a good fit to the data \cite{decayb}.  Similarly, decoherence
of the neutrino beam predicts
\begin{eqnarray}
P_{\mu\mu}=1-\sin^22\theta\left(1-e^{-\gamma L}\right),\nonumber
\end{eqnarray}
and this scenario has been shown \cite{lisi}
to give a good fit to the data.

Before the announcement against sterile oscillations in
both solar \cite{ysuzuki} and atmospheric \cite{SKs}
neutrino data by the Superkamiokande group
in June 2000, several groups \cite{extra}
claimed that scenarios
of large extra dimension give a good fit to the data of
solar neutrinos or atmospheric neutrinos.  However,
oscillations predicted by those scenarios are basically
sterile oscillations and they may no longer give a good fit to the data.

The behaviors of the surviving probability $P(\nu_\mu\rightarrow\nu_\mu)$
in vacuum is plotted as a function of $L/E$ in Fig. \ref{fig:p}
(taken from Ref.~\cite{pakvasa}) for various scenarios.
The main difference between the oscillation due to mass and the
exotic scenarios is that the former has
dip in the surviving probability and it will be possible to check
the existence of the dip in
long baseline experiments or atmospheric neutrino experiments
like MONOLITH \cite{monolith} in the future.

\section{Summary}

In this talk I have reviewed various solutions of the atmospheric
neutrino data.  The two flavor $\nu_\mu\leftrightarrow\nu_\tau$
oscillation with almost maximal mixing gives an excellent fit to the
data, and consequently so does the three flavor oscillation with small
$\theta_{13}$.  As for four flavor oscillation scenarios, there are
two types of schemes.  The (2+2)-scheme is still consistent with both
the solar and atmospheric neutrino data, since both solar and
atmospheric neutrino oscillations are hybrid of active and sterile
oscillations.  The (3+1)-scheme is allowed for $\Delta
m^2_{\mbox{\rm{\scriptsize LSND}}}$ =0.9, 1.7, 6.0 eV$^2$ and in this
scheme the solar neutrino deficit is accounted for by active
oscillations while the atmospheric neutrino anomaly is explained by
hybrid of active and sterile oscillations.
There are also a couple of exotic scenarios which give a good fit
to the data.  They are scenarios of neutrino decay and decoherence,
and these hypotheses can be checked by looking at the oscillation
dip in the probability in the future experiments.

\section*{Acknowledgments}
I would like to thank the organizers for invitation.  This
research was supported in part by a Grant-in-Aid for Scientific
Research of the Ministry of Education, Science and Culture,
\#12047222, \#10640280.


\section*{References}
\begin{thebibliography}{99}

\bibitem {Kamatm}
K.S. Hirata et al.,
Phys. Lett. {\b f B205} (1988) 416; Phys. Lett. {\bf B280} (1992) 146;
Y. Fukuda et al., Phys. Lett. {\bf B335} (1994) 237. 

\bibitem {Kamup}
S. Hatakeyama et al.,  
Phys. Rev. Lett. {\bf 81}, 2016 (1998).

\bibitem{IMB}
D. Casper et al., Phys. Rev. Lett. {\bf 66} (1989) 2561;
R. Becker-Szendy et al., Phys. Rev. {\bf D46}, 3720 (1992).

\bibitem{SKatm}
Y. Fukuda et al., Phys. Lett. {\bf B433}, 9 (1998);
Phys. Lett. {\bf B436}, 33 (1998);
Phys. Rev. Lett. {\bf 81}, 1562 (1998).

\bibitem {SKup}
Y. Fukuda et al.,
Phys. Rev. Lett. {\bf 82}, 2644 (1999).

\bibitem{SKs}
Y. Fukuda et al., Phys. Rev. Lett. {\bf 85}, 3999 (2000).

\bibitem{learned}
J.G. Learned, hep-ex/0007056.

\bibitem{mcgrew}
\fussy
C. McGrew's talk, in these proceedings
(http://www-sk.icrr.u-tokyo.ac.jp/
noon/2/transparency/1207/08/index1.html).

\bibitem {soudan2}
W.W.M. Allison et al., Phys. Lett. {\bf B449}, 137 (1999).

\bibitem {macro}
M. Spurio, Nucl. Phys. (Proc. Suppl.) {\bf 85} 37 (2000).

\bibitem{atmth}
J.G. Learned, S. Pakvasa, and T.J. Weiler, 
Phys. Lett. {\bf B207}, 79 (1988);
V. Barger and K. Whisnant, 
Phys. Lett. B {\bf 209}, 365 (1988);
K. Hidaka, M. Honda and S. Midorikawa,
Phys. Rev. Lett. {\bf 61}, 1537 (1988).

\bibitem{atmanalysis}
O. Yasuda, hep-ph/9602342; hep-ph/9706546; hep-ph/9809205;
Phys. Rev. {\bf D58}, 091301 (1998);
Nucl. Phys. B (Proc. Suppl.) {\bf 77}, 146 (1999);
Acta Phys. Pol. {\bf B30}, 3089 (1999);
R. Foot, R.R. Volkas and O. Yasuda, Phys. Rev. {\bf D57}, 1345
(1998); Phys. Lett. {\bf B421}, 245 (1998):
Phys. Rev. {\bf D58}, 13006 (1998); Phys. Lett. {\bf B433}, 82 (1998);
R. Foot, C.N. Leung and O. Yasuda, Phys. Lett. {\bf B443}, 185 (1998);
G.L. Fogli, E. Lisi, D. Montanino and G. Scioscia,
Phys. Rev. {\bf D55}, 4385 (1997);
G.L. Fogli, E. Lisi, A. Marrone and D. Montanino,
Phys. Lett. {\bf B425}, 341 (1998);
G.L. Fogli, E. Lisi and A. Marrone
Phys. Rev. {\bf D57}, 5893 (1998);
G.L. Fogli, E. Lisi, A. Marrone and G. Scioscia,
Phys. Rev. {\bf D59}, 033001 (1999);
Phys. Rev. {\bf D59}, 117303, (1999);
Phys. Rev. {\bf D60}, 053006, (1999);
Nucl. Phys. B (Proc. Suppl.) {\bf 85}, 159 (2000);
J.W. Flanagan, J.G. Learned and S. Pakvasa, 
Phys. Rev. {\bf D57}, 2649 (1998);
M.C. Gonzalez-Garcia, H. Nunokawa, O.L.G. Peres, T. Stanev
and J.W.F. Valle, Phys. Rev. {\bf D58}, 033004 (1998);
M.C. Gonzalez-Garcia, M.M. Guzzo, P.I. Krastev
H. Nunokawa, O.L.G. Peres, V. Pleitez,
J.W.F. Valle and  R. Zukanovich Funchal,
Phys. Rev. Lett. {\bf 82}, 3202 (1999);
M.C. Gonzalez-Garcia, H. Nunokawa, O.L.G. Peres
and J.W.F. Valle, Nucl. Phys. {\bf B543}, 3 (1999);
N. Fornengo, M.C. Gonzalez-Garcia and J.W.F. Valle,
JHEP {\bf 0007}, 006 (2000); Nucl. Phys. {\bf B580}, 58 (2000);
O.L.G. Peres, A.Yu. Smirnov
Phys. Lett. {\bf B456}, 204-213,1999;
E. Akhmedov, P. Lipari and M. Lusignoli,
Phys. Lett. {\bf B300}, 128 (1993);
P. Lipari and M. Lusignoli,
Phys. Rev. {\bf D57}, 3842 (1998);
Phys. Rev. {\bf D60}, 013003 (1999);
J.G. Learned, S. Pakvasa and J.L. Stone,
Phys. Lett. {\bf B435}, 131 (1998);
Q.Y. Liu and A.Yu. Smirnov,
Nucl. Phys. {\bf B525}, 505 (1998);
Q.Y. Liu, S.P. Mikheyev and A.Yu. Smirnov,
Phys. Lett. {\bf B440}, 319 (1999).
E. Kh. Akhmedov, A. Dighe, P. Lipari, A. Yu. Smirnov,
Nucl. Phys. {\bf B542}, 3 (1999);
S. Choubey and S. Goswami, Astropart. Phys. {\bf 14}, 67 (2000);
S. Choubey, S. Goswami and K. Kar, hep-ph/0004100;
T. Teshima, T. Sakai, O. Inagaki,
Int. J. Mod. Phys. {\bf A 14}, 1953 (1999);
T. Teshima, T. Sakai, Prog. Theor. Phys. {\bf 101}, 147 (1999);
Prog. Theor. Phys. {\bf 102}, 629 (1999);
Phys. Rev. {\bf D62}, 113010 (2000).
\bibitem{homestake}
B.T. Cleveland et al., Nucl. Phys. B (Proc. Suppl.) {\bf 38}, 47 (1995).

\bibitem{Kamsol}
Y. Fukuda et al., Phys. Rev. Lett. {\bf 77}, 1683 (1996)
and references therein.

\bibitem{SKsol}
Y. Suzuki, Nucl. Phys. B (Proc. Suppl.) {\bf 77}, 35 (1999)
and references therein.

\bibitem{ysuzuki}
Y. Suzuki, talk at {\em 19th International Conference on Neutrino Physics
and Astrophysics} (Neutrino 2000), Sudbury, Canada, June 16-22, 2000\\
(http://nu2000.sno.laurentian.ca/ Y.Suzuki/).

\bibitem{sage}
V.N. Gavrin, Nucl. Phys. B (Proc. Suppl.) {\bf 77}, 20 (1999)
and references therein.

\bibitem{gallex}
T.A. Kirsten, Nucl. Phys. B (Proc. Suppl.) {\bf 77}, 26 (1999)
and references therein.

\bibitem{lsnd}
C. Athanassopoulos {\em et al}.,
Phys. Rev. Lett. {\bf 77}, 3082 (1996);
Phys. Rev. C {\bf 54}, 2685 (1996);
Phys. Rev. Lett. {\bf 81}, 1774 (1998);
Phys. Rev. C {\bf 58}, 2489 (1998);
D.H. White, Nucl. Phys. Proc. Suppl. {\bf 77}, 207 (1999).

\bibitem{mills}
G. Mills, talk at {\em 19th International Conference on Neutrino Physics
and Astrophysics} (Neutrino 2000), Sudbury, Canada, June 16--22, 2000\\
\fussy
(http://nu2000.sno.laurentian.ca/G.Mills/).

\bibitem{chooz}
M. Apollonio et al., Phys. Lett. {\bf B420}, 397 (1998);
Phys. Lett. {\bf B466}, 415 (1998).

\bibitem{3to2}
C.-S. Lim, Proc. of the
BNL Neutrino Workshop on Opportunities for Neutrino Physics at BNL,
Upton, N.Y., February 5-7, 1987, ed. by M. J. Murtagh, p111;
A. Yu. Smirnov, Proc. of the Int Symposium on
Neutrino Astrophysics, Takayama/Kamioka 19 - 22 October 1992,
ed. by Y. Suzuki and K. Nakamura, p.105.

\bibitem{flmm}
G.L. Fogli, E. Lisi, A. Marrone and D. Montanino, hep-ph/0009269.

\bibitem{gmpv}
M.C. Gonzalez-Garcia, M. Maltoni, C. Pena-Garay and J.W.F. Valle,
Phys. Rev. {\bf D63} (2001) 033005.

\bibitem{y1}
O. Yasuda, hep-ph/0006319.

\bibitem{flm}
G.L. Fogli, E. Lisi and A. Marrone,
Phys. Rev. {\bf D63}, 053008 (2001).

\bibitem{cdhsw}
F. Dydak {\em et al}., Phys. Lett. B {\bf 134}, 
 281 (1984).

\bibitem{pdg}
Review of Particle Physics, Particle Data Group,
Eur. Phys. J. {\bf C3}, 1 (1998).

\bibitem{oy}
 N. Okada and O. Yasuda, Int. J. Mod. Phys. {\bf A 12}, 3669 (1997).

\bibitem{bgg}
 S.M. Bilenky, C. Giunti and W. Grimus, hep-ph/9609343;
Eur. Phys. J. {\bf C1}, 247 (1998).

\bibitem{bugey}
B. Ackar et al., Nucl. Phys. {\bf B434}, 503 (1995).

\bibitem{bklw}
V. Barger, B. Kayser, J. Learned, T. Weiler and K. Whisnant,
Phys. Lett. {\bf B489}, 345 (2000).

\bibitem{y2}
O. Yasuda, unpublished.

\bibitem{vep}
M. Gasperini, Phys. Rev. {\bf D38}, 2635 (1988);
A. Halprin and C. N. Leung, Phys. Rev. Lett. {\bf 67}, 1833 (1991)
\bibitem{lorentz}
S.~Coleman and S.L.~Glashow, Phys. Lett. {\bf B405}, 249 (1997);
D.~Colladay and V.A.~Kostelecky, Phys. Rev. {\bf D55}, 6760 (1997).
\bibitem{torsion} V. De Sabbata and M. Gasperini, Nuovo Cim. 65 {\bf A} (1981)
479.
\bibitem{FCNC}
E. Roulet, Phys. Rev. D {\bf 44} (1991) 935; M. M. Guzzo, 
A. Masiero and S. T. Petcov, Phys. Lett. B {\bf 260} (1991) 
154; V. Barger, R. J. N. Phillips and K. Whisnant, Phys. 
Rev. D {\bf 44} (1991) 1629.
\bibitem{decaya}
V. Barger, J.G. Learned, S. Pakvasa and T.J. Weiler,
Phys. Rev. Lett. {\bf 82}, 2640 (1999);
P. Lipari and M. Lusignoli, Phys. Rev. {\bf D60}, 013003 (1999);
G.L. Fogli, E. Lisi and A. Marrone, Phys. Rev. {\bf D59}, 117303 (1999);
S. Choubey and S. Goswami, Astropart. Phys. {\bf 14}, 67 (2000).
\bibitem{decayb}
V. Barger, J.G. Learned, P. Lipari, M. Lusignoli,
S. Pakvasa and T.J. Weiler, Phys. Lett. {\bf B462}, 109 (1999).
\bibitem{lisi}
E. Lisi, A. Marrone and D. Montanino,
Phys. Rev. Lett. {\bf 85}, 1166 (2000);
\bibitem{extra}
R.N. Mohapatra, S. Nandi and A. Perez-Lorenzana,
Phys. Lett. {\bf B466}, 115 (1999);
R. N. Mohapatra and A. Perez-Lorenzana, Nucl. Phys. {\bf B576}, 466 (2000);
Y. Grossman and M. Neubert, Phys. Lett. {\bf B474}, 361 (2000);
G. Dvali and A. Yu. Smirnov, Nucl. Phys. {\bf B563}, 63 (1999);
R. Barbieri, P. Creminelli and A. Strumia,
Nucl. Phys. {\bf B585}, 28 (2000).

\bibitem{pakvasa}
S. Pakvasa, hep-ph/0008193.

\bibitem{monolith}
A. Geiser, hep-ex/0008067; P. Antonioli, hep-ex/0101040.

\end{thebibliography}
\end{document}